# Dynamic fabrication method of SNAP microresonators

ZIJIE WANG,[1,2] MANUEL CRESPO-BALLESTEROS,[1] CHRISTOPHER COLLINS,[1] YONG YANG, [2] QI ZHANG, [2] XIAOBEI ZHANG, [2] AND MICHAEL SUMETSKY [1*]

[1] Aston Institute of Photonics Technology, Aston University, Aston Triangle, Birmingham, UK, B47ET
[2] Key Laboratory of Specialty Fiber Optics and Optical Access Networks, Joint International Research Laboratory of Specialty Fiber Optics and Advanced Communication, Shanghai Institute for Advanced Communication and Data Science, Shanghai University, Shanghai 200444, China
*m.sumetsky@aston.ac.uk

**Surface Nanoscale Axial Photonics (SNAP) technology has demonstrated the record subangstrom fabrication precision of optical microresonators and resonant photonic circuits at the optical fiber surface. However, fabrication errors arising from fluctuations of temperature, inscription parameters, alignment inconsistencies, and other factors did not allow researchers to achieve the subangstrom precision without sophisticated postprocessing. Here we show that the key fabrication method of SNAP structures – $CO_2$ laser beam optical fiber annealing – suffers from significant fiber displacements which may introduce a few percent fabrication errors. To suppress the effects of misalignment, we develop a dynamic fabrication method employing a translating beam exposure and demonstrate its excellent precision. The effective fiber radius variation of ~ 10 nm is introduced with an error of ~ 0.1 angstrom. We suggest that the remaining fabrication errors can be attributed to laser power fluctuations.**

Microresonators serve as fundamental building blocks in the development of miniature photonic circuits and microdevices, playing a key role in diverse applications of modern photonics [1-6]. The ultraprecise fabrication of individual and coupled microresonators is essential for enabling practical implementations of microwave photonics [7, 8], optical frequency comb generation [9, 10], classical and quantum information processing [11], biosensing [12], and resonant optomechanical systems [13, 14]. However, it has become evident over the past years that a remarkable *few nanometers fabrication precision* of optical microresonators achieved in microphotonics technologies [15-17] remains insufficient to fully exploit the potential of microresonators in these applications. For several important cases, the required precision has to reach a *picometer scale* [18-20]. Despite the recent significant progress towards improving the fabrication precision of microphotonic devices beyond a few nanometers, which was primarily based on their postprocessing (see [18-22] and references therein) the research towards picometer fabrication precision is currently in its initial stage.

Surface Nanoscale Axial Photonics (SNAP) offers a promising solution to this challenge by enabling the fabrication of microresonators with sub-angstrom precision and ultralow loss at the optical fiber surface [18, 21-32]. In the SNAP platform, a nanoscale effective radius variation (ERV) is introduced along the optical fiber leading to localized changes in its cutoff wavelengths (CWLs) that govern the propagation of whispering gallery modes (WGMs) adjacent the fiber surface. The nanoscale ERV provides unprecedented control over WGMs, allowing them to circulate around the optical fiber surface while propagating slowly along the fiber axis.

Various techniques have been developed to fabricate SNAP structures, each exploring different physical mechanisms to introduce and control ERVs. These methods include annealing using $CO_2$ laser [21, 22], flame [23], and electric arc [24], femtosecond laser inscription [25, 26, 27], microfluidic-assisted slow processing (coined as the slow cooking of optical microresonators) [28], as well as introduction of tunable SNAP resonators by local fiber heating [29, 30], fiber bending [31, 32, 33], and side-coupled optical fiber coupling [34]. The precise engineering of SNAP devices using these methods enables their unique applications in next generation telecommunications, computing, and sensing technologies.

Among these fabrication techniques of SNAP structures, $CO_2$ laser annealing is prominent due to its superior precision, simplicity, and robustness [18, 21, 22]. This method typically achieves the fabrication precision on the order of a few angstroms, while sub-angstrom precision is attainable through multiple post-processing iterations [21, 22]. The need for iterative refinement arises due to inherent fluctuations in laser power, temperature variations, nonuniformities in the original optical fiber, and imperfections in system alignment. For example, a structure consisting of 30 coupled SNAP microresonators was initially fabricated with a precision of 6 Å using $CO_2$ laser annealing, which was subsequently improved to 0.7 Å through multiple post-processing steps [21]. Similarly, in [22], two coupled microresonators were fabricated with an ultrahigh precision of 0.17 Å after multiple optimization cycles. To our knowledge, the fabrication precision achieved in Refs. [21, 22] remains the highest compared to more recent demonstrations (see [19, 20] and references therein).

To develop a fabrication method of SNAP devices with subangstrom and, eventually, picometer precision without or with minimum possible post-processing steps, it is critical to clarify and suppress the physical effects leading to the noted fabrication errors.

Here we address this challenge for the $CO_2$ laser annealing fabrication method and show that laser beam heating may introduce significant displacement of an optical fiber leading to fluctuations of the annealing process. To suppress this and other misalignment effects, we develop the method of *translational exposure*. In contrast to the stationary beam exposure employed previously [18, 21, 22], the annealing effect in our new approach is determined by the speed of the laser beam crossing the fiber. Consequently, the fiber misalignment in the plane normal to this translation is compensated, since then the beam will expose the slowly moving fiber independently of its changing position. At the same time, the heating and cooling processes become smooth in time and do not introduce fiber vibrations. While, in the experiments presented below, the developed approach does not improve the fabrication precision significantly, this study allows us to suggest that the remaining fabrication errors can be attributed to laser power fluctuations.

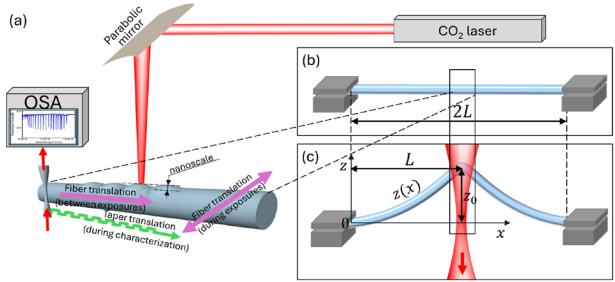

Fig. 1. (a) Illustration of the previously developed method for the fabrication of SNAP microresonators using $CO_2$ laser annealing. (b) A straight optical fiber with length $2L$ clamped at the ends. (c) The same fiber bent due to local heating in the middle.

Figure 1(a) illustrates the $CO_2$ laser annealing fabrication method of SNAP structures developed previously (see [18, 21, 22] and references therein). In this method, the local annealing of an optical fiber with a stationary laser beam, which is focused with a parabolic cylinder mirror, creates SNAP microresonators with nanoscale ERV. The parabolic mirror focuses the beam along the exposed fiber axis direction and excludes beam's focusing along the direction transverse to the fiber. Consequently, in contrast to a spherical mirror, a parabolic mirror allows us to smooth the exposure process and avoid rapid fiber heating that may lead to vibrations described by Eq. (4). The ERV is introduced by relaxing the local residual stress frozen in during the optical fiber drawing process. The fabricated microresonator structures are then characterized by measuring their transmission spectrum using a coupled transverse microfiber taper connected to an optical spectrum analyzer (OSA) and scanned along the fiber as shown in Fig. 1(a). Multiple transmission spectra successively measured along the fiber axis with required resolution are collected in a spectrogram [18, 21-34].

Here we note that heating the originally straight optical fiber clamped at the ends (Fig. 1(b)) can lead to its bending as illustrated in Fig. 1(c). To estimate this effect, we suggest that, during the annealing process, the fiber temperature is increased by $\Delta T \sim 1000$ K along $\Delta L_{heat} \sim 0.3$ mm of the fiber length [29]. This leads to the fiber expansion $\Delta L = \alpha \Delta T \Delta L_{heat} \sim 0.2$ μm, where the value of thermal expansion coefficient for silica is set to $\alpha = 0.5 \cdot 10^{-6}$ K$^{-1}$. Assuming the original length of the clamped fiber equal to $2L$, and the maximum shift along the laser beam axis $z$ equal to $z_0$ (Fig. 1(c)), we find the deviation of the fiber axis as a function of the original axial coordinate $x$ using the Euler-Bernoulli beam theory [35], which yields for $z \leq L$:

$$z(x) = \frac{z_0}{2}\left[3\left(\frac{x}{L}\right)^2 - \left(\frac{x}{L}\right)^3\right]. \quad (1)$$

As shown in Fig. 1(c), we assume that the fiber is bent towards the incoming laser beam since the fiber heats stronger at the part of its surface facing the beam [36]. Calculating the fiber axial length from Eq. (1) under the assumption $\Delta L \ll L$, we find $\Delta L = 6z_0^2/5L$ and, thus,

$$z_0 = \sqrt{\frac{5}{6}\Delta L \cdot L}. \quad (2)$$

In our experiment considered below, we have $L = 20$ mm, so that for $\Delta L \sim 0.2$ μm found above we have 300 times larger $z_0 \sim 60$ μm found from Eq. (2). This deviation can introduce a variation of the fiber heating power noticeably affecting the fabrication precision. Indeed, we find for the laser beam intensity distribution near its focus [37]:

$$I(x,z) = \frac{I_0 w_0}{w(z)} \exp\left(-\frac{2x^2}{w(z)^2}\right),$$
$$w(z) = w_0\sqrt{1+\frac{z^2}{z_R^2}}, \quad w_0 = \frac{\lambda f}{\pi w_{inc}}, \quad z_R = \frac{\pi w_0^2}{\lambda}. \quad (3)$$

Here $f$ is the parabolic mirror focal length, $\lambda$ is the laser wavelength, $w_{inc}$ and $w_0$ are the incident and the focused laser beam radii, and the coordinate $z = 0$ (the original position of the optical fiber, see Figs. 1(b) and (c)) is assumed to coincide with the beam focus. From Eq. (3), at $\lambda \sim 10$ μm, $z = z_0 \sim 60$ μm, and the focal length $f = 15$ mm considered in our experiment below, we find $I(0, z_0) = 0.98 I_0$, which corresponds to the laser beam intensity variation ~ 2%. This variation is comparable with the power fluctuations ~ 5% of the Synrad 48 $CO_2$ laser used in our experiments.

Additionally, rapid fiber heating can excite its bending vibrations. From the Euler-Bernoulli beam theory [38], we have for the fundamental bending frequency of the fiber:

$$v_1 \simeq 0.9\frac{r_0}{L^2}\sqrt{\frac{E}{\rho}}. \quad (4)$$

Here $E$ is the Young modulus, $\rho$ is the density, and $r_0$ is the radius of the fiber. For a silica fiber, setting $E = 7.2 \cdot 10^{10}$ Pa, $\rho = 2200$ kg/m³, $r_0 = 19$ μm, and $L = 20$ mm, we find from

Eq. (4) $\nu_1 \cong 245$ Hz. We suggest that excitation of fiber vibrations with such frequency and amplitude of the order of $z_0 \sim 60$ μm found above can introduce noticeable fluctuations into the fiber annealing process affecting the fabrication precision.

It follows from Eqs. (2), (3) and (4) that the relative laser heating intensity deviation $(I_0 - I(0, z_0))/I_0$ increases and the fiber bending frequency $\nu_1$ rapidly decreases with growing of the clamped fiber length $2L$. The plots of these parameters as a function of the fiber length are shown in Fig. 2. It is seen that, for the 10 cm fiber length, the intensity deviation can reach 4%, while the fiber vibration frequency can be as small as 40 Hz.

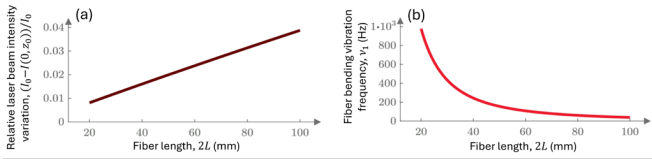

**Fig. 2.** (a) The relative laser intensity deviation from its focal maximum, $(I_0 - I(0, z_0))/I_0$, as a function of the fiber length $2L$. (b) Fundamental fiber bending frequency $\nu_1$ as a function of the fiber length $2L$.

Here we develop a method to suppress the effects of the fiber misalignment described above as well as to compensate for the possible original fiber misalignment in the plane normal to the laser beam axis. Instead of heating the fiber with laser beam power rapidly switched on and then held over a predetermined (typically, a few seconds) period (Fig. 1(a)) employed previously [21, 22], we translate the fiber normally to the beam, so that the fiber crosses the beam for a certain period of time as illustrated in Fig. 3(a). If the crossing speed is small enough then the fiber temperature increases and then decreases smoothly in time. In our experiment described below, the beam power and the speed of translation are also optimized to arrive at the required nanoscale ERV.

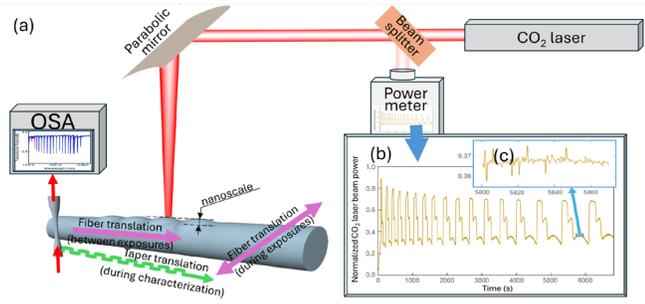

**Fig. 3.** (a) Illustration of the developed method for the fabrication of SNAP microresonators using $CO_2$ laser annealing. (b) Time-dependent variation of laser power measured by a power meter for the characteristic output laser power ~ 1 W. (c) Magnified section outlined by the blue rectangle in plot (b) corresponding to close to permanent behavior of laser power over one minute time.

As follows from our calculations above, the characteristic heating time during the ERV inscription process should be much greater than the characteristic period of fiber vibrations $\sim 1/\nu_1$. In the absence of laser beam power fluctuations, this condition ensures a smooth-in-time and reproducible fiber heating process.

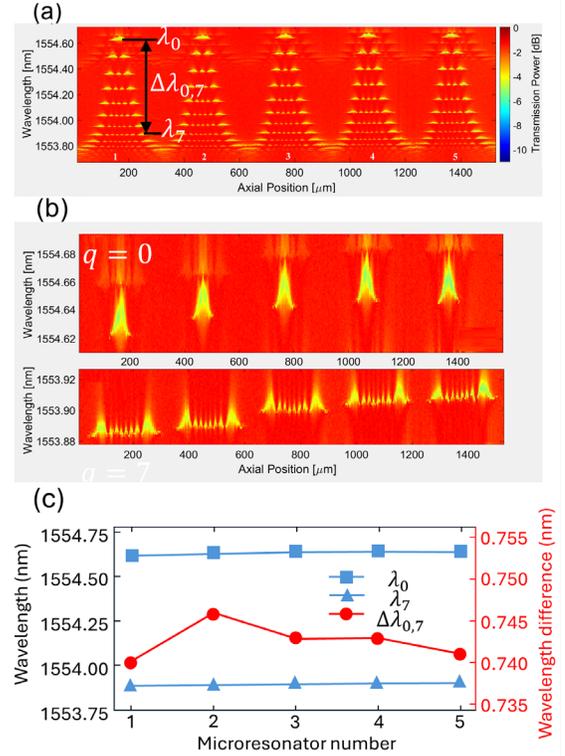

**Fig. 4.** (a) Spectrogram of five SNAP microresonators. (b) Transmission spectra at axial positions of $z_1$=122 μm, $z_2$=202 μm, $z_3$=422 μm, $z_4$=502 μm, $z_5$=722 μm, $z_6$=802 μm. (c) Enlarged view of the resonance wavelengths at points A-F.

A characteristic laser power variation in time for a Synrad 48 laser $CO_2$ laser used in our experiments and measured with a Coherent power meter is shown in Fig. 3(b). It is seen that, in the open loop regime considered, the laser power (set to a relatively low value of ~ 1 W to enable slow heating beam translation) exhibited characteristic slow in time large power variations and also rapid power fluctuations of ~ 2% and slightly less rapid of ~ 5% visible in the magnified segment of this plot in Fig. 3(c). Figures 3(b) and (c) show that the time durations when the laser power remains constant with ~ 5% fluctuations is close to one minute. In our experiments, the laser power was continuously controlled to ensure that its power variation does not go beyond the fluctuations shown in Fig. 3(c). Since the characteristic exposure of a single SNAP microresonator does not exceed a few seconds, we were able to create multiple microresonators during a single stable power period.

We demonstrate the proposed method by fabricating five SNAP microresonators on a silica fiber with radius $r_0 = 19$ μm. To ensure the fiber straight alignment without introducing stress, we dropped a ~ 8 cm long fiber segment, uncoated in the middle, onto a substrate of two parallel and aligned plates with 4 cm gap and then glued its ends to the substrate. To ensure a one-second characteristic translation time of the fiber through the focused laser beam, which satisfies the required slowness of heating and cooling processes, and to enable the required value of inscribed ERV, the laser power was set to ~ 1 W, and the transverse translation fiber speed was set to 0.6 mm/s. For comparison, we found that a similar ERV can be inscribed in a silica fiber with radius $r_0 = 62.5$ μm (not considered in this brief report) by reducing the translation speed to ~ 0.1 mm/s. The inscription of these resonators was performed during a single stabilized laser period outlined in Fig. 3(c).

Fig. 4(a) shows the spectrogram of fabricated microresonators measured by scanning the fiber with a transverse microfiber taper with steps of 2 μm along the fiber axis, as illustrated in Figs. 1(a) and 3(a). The spectral range in this figure was chosen to show the variation of a single cutoff wavelength $\lambda_c(x)$ along the fiber axis $x$, with the original value close to 1553.8 nm. The introduced microresonators have ~ 240 μm length along the fiber axis. Their cutoff wavelength variation (CWV) $\Delta\lambda_c(x)$ has the amplitude of $\Delta\lambda_{c0} \cong 0.8$ nm, which corresponds to the fiber ERV of $\Delta r_{eff} = r_0 \Delta\lambda_{c0}/\lambda_c \cong 10$ nm.

For each of the fabricated microresonators, the CWV shown in the spectrogram confines 10 modes with axial quantum numbers $q = 0,1,...,10$. Sections of this spectrogram magnified along the wavelength axis near eigenwavelengths with $q = 0$ and $q = 7$ are shown in Figs. 4(b). It is seen from these figures that the microresonators are not exactly equal having the CWV $\lambda_c(x)$ difference of ~ 0.03 nm corresponding to the ERV difference of ~ 4 angstroms. We notice that this fabrication precision does not take into account the original ERV of the optical fiber, which is typically less than an angstrom along the millimeter fiber length (see, e.g., Fig. 13 in Ref. [18]).

To characterize the actual fabrication precision of the introduced microresonators, we eliminate the original fiber radius nonuniformity by comparing the difference $\Delta\lambda_{0,7}$ between microresonators' eigenwavelengths $\lambda_0$ and $\lambda_7$ with quantum numbers $q = 0$ and $q = 7$ plotted in Fig. 4(c). We find from this comparison that the variation of this difference having an approximate value of $\Delta\lambda_{0,7} \cong 0.8$ nm is ~ 0.007 nm corresponding to the ~ 1% ERV inscription error. The relative ERV error, $\delta r_{eff}/\Delta r_{eff}$, caused by the relative power fluctuations, $\delta P/P$, can be estimated from the relation $\delta\lambda_{0,7}/\Delta\lambda_{0,7} = \delta r_{eff}/\Delta r_{eff} \cong \delta P/P$. A few percent fabrication error, fairly well correlated with a few percent power fluctuations of the $CO_2$ laser beam, was observed in our other similar experiments. This allows us to suggest that the remaining fabrication errors can be attributed to the laser power fluctuations. These fluctuations can be suppressed by advanced methods of power stabilization [39] as well as by setting the laser power to a value at which the fractional power fluctuations are minimized, then attenuating the power down to the required values.

**Funding.** National Natural Science Foundation of China grants 62205192 and 62405175, China Scholarship Council grant 202306890079, Leverhulme Trust grant RPG-2022-014, Engineering and Physical Sciences Research Council (EPSRC) grants EP/W002868/1 and EP/X03772X/1.

**Disclosures.** The authors declare no conflicts of interest.

**Data availability.** Data underlying the results presented in this paper may be obtained from the corresponding author upon reasonable request.